\documentclass[reprint,onecolumn,notitlepage,superscriptaddress,amsmath,amssymb,aps,prb,showpacs]{revtex4-1}
\usepackage[]{graphicx,subfigure}
\usepackage{dcolumn}
\usepackage{bm}
\usepackage{hyperref}
\usepackage{multirow}
\usepackage{tabularx}
\usepackage{siunitx}

\usepackage[utf8]{inputenc}
\usepackage{wrapfig}

\begin{document}

\title{High-quality Blazed Gratings through Synergy
between E-Beam lithography and Robust Characterisation Techniques}

\author{Analía F. Herrero}
 \email[Corresponding author: ]{analia.fernandez.herrero@ptb.de}
 \affiliation{Helmholtz-Zentrum Berlin, Albert-Einstein Str. 15, 12489 Berlin, Germany}
 \affiliation{Physikalisch-Technische Bundesanstalt, Abbestr. 2-12, 10587 Berlin, Germany}

\author{Nazanin Samadi}%
\affiliation{Paul Scherrer Institut, Center for Photon Science, Forschungsstrasse 111, 5232 Villingen PSI, Switzerland}
\affiliation{Deutsches Elektronen-Synchrotron DESY, Notkestr. 85, 22607 Hamburg, Germany}

\author{Andrey Sokolov}
 \affiliation{Helmholtz-Zentrum Berlin, Albert-Einstein Str. 15, 12489 Berlin, Germany}
\author{Grzegorz Gwalt}
 \affiliation{Helmholtz-Zentrum Berlin, Albert-Einstein Str. 15, 12489 Berlin, Germany}
\author{Stefan Rehbein}
 \affiliation{Helmholtz-Zentrum Berlin, Albert-Einstein Str. 15, 12489 Berlin, Germany}

\author{Anke Teichert}
\affiliation{NOB - Nano Optics Berlin, Krumme Strasse 64, 10625 Berlin, Germany}

\author{Bas Ketelaars}
\affiliation{RAITH B.V., De Dintel 27-A, 5684 PS Best, The Netherlands}

\author{Christian Zonnevylle}
\affiliation{RAITH B.V., De Dintel 27-A, 5684 PS Best, The Netherlands}

\author{Thomas Krist}
\affiliation{NOB - Nano Optics Berlin, Krumme Strasse 64, 10625 Berlin, Germany}

\author{Christian David}%
\affiliation{Paul Scherrer Institut, Center for Photon Science, Forschungsstrasse 111, 5232 Villingen PSI, Switzerland}

\author{Frank Siewert}
 \affiliation{Helmholtz-Zentrum Berlin, Albert-Einstein Str. 15, 12489 Berlin, Germany}

\begin{abstract}

Maintaining the highest quality and output of photon science in the VUV-, EUV-, soft- and tender-X-ray energy ranges requires high-quality blazed profile gratings. Currently, their availability is critical due to technological challenges and limited manufacturing resources. In this work we discuss the opportunity of an alternative method to manufacture blazed gratings by means of electron-beam lithography (EBL). We investigate the different parameters influencing the optical performance of blazed profile gratings produced by EBL and develop a robust process for the manufacturing of high-quality blazed gratings using polymethyl methacrylate (PMMA) as high resolution, positive tone resist and ion beam etching.

\end{abstract}
\maketitle

\section{Introduction}

The EUV, VUV and soft X-ray energy regions are of great interest for industry and academia alike. The so-called water window, essential in natural sciences, is located within these regions, for instance. The EUV region, in turn, has seen a great development in materials science and coatings for the semiconductor industry and astrophysics during the last decades. On the other hand, accelerator-based photon sources have seen an enormous improvement in terms of brilliance, stability, and coherence \cite{Eriksson_DLSR}. Not only large-scale facilities but also lab sources have increased their stability and power in the EUV and soft X-ray energy ranges. For photonic sciences to capitalize on these advances, the availability of high-quality blazed gratings is essential~\cite{underwood_ml-echelle-grating_1995,Lin_MLgrating_euv_2008,Cocco_2022_wavefront}. Blazed gratings show substantially higher diffraction efficiency than their laminar counterparts. They are used at monochromators in broadband emission sources as well as analyzers in spectroscopical applications at those energy regions. In case of applications at Free Electron Lasers blazed gratings are of interest as they show a higher damage threshold compared to lamellar gratings\cite{Gaudin_2012_investigation, Krzywinski_2018_damage}. Furthermore, it has been shown how the tender X-ray region can also benefit from the availability of high-quality blazed gratings coated with a dedicated multilayer (ML)\cite{Sokolov:19,Werner_2023,wen_sokolov_tener_2024}. Working at photon energies between 1.5 keV and 3keV usually need to operate crystal monochromators near to normal incidence, which causes highly thermal instabilities and limits their usability. However, using multilayer coated blazed gratings, efficiencies up to 60 \% can be achieved \cite{Sokolov:19}.  Thus, maintaining the highest quality of beamline and end-station performance at photon sources from the VUV to the tender X-ray region requires research into new and more versatile methods of producing such gratings. 

Mechanical ruling of blazed gratings is a common and established technique since a long time \cite{siewert_2018,Dziarzhytski_2028_spectrometer}. However, it is time consuming, of high technological risk and with only very few places around the world offering it. Moreover, the method is less versatile than others that can locally adapt the profile to counteract any feature of the structure or substrate. During the last decades, other methods have been investigated. Reasonable results have been obtained for low line densities and very shallow angles by nanoimprinting~\cite{Voronov:23}. However, high-resolution spectrometers require high line densities, challenging for most of the above mentioned methods. Also high blaze angles at high line density gratings are a challenge in imprinting technology. First tests on anisotropic etching through a mask or patterned area following by anisotropic etching for the production of a blazed profile grating date back to the 1980s ~\cite{fujii_1980_optical,Philippe_1985_echelette}. This method allows a very low facet roughness but results on a plateau on the apex region of the grating that negatively impacts the grating efficiency. Additionally steps have been included to reduce the area of this region while maintaining the region sharp \cite{nie_2020_anisotropic,zha_design_2022, Voronov:22}. However, the control of the blaze to antiblaze angle cannot be achieved and the efficiency is thus, affected. Also, this approach is limited to plane substrates while curved (spherical or cylindrical) substrates cannot be patterned with sufficient precision. Grey-tone electron-beam lithography (EBL) has also been explored in the last decades as an alternative method for the production of optical elements \cite{henke_simulation_1994,zhitian_towards_2020}. Tests for the manufacturing of blazed profile structures have been conducted using dose variation of the EBL ~\cite{Schleunitz_2010_dose-variation} as well as with frequency variation~\cite{mccoy_2028_fabrication} of the e-beam writer exposure. In this method, the resist is exposed to a variable dose, and afterwards, the patterned resist is transferred into the substrate, for instance by ion beam etching \cite{BORZENKO1994337} or oxidation \cite{patent_oxidation_cd} depending on the resist. As the slope of the facet can only be approximated by a certain number of steps, methods have been developed to reduce the waviness of the facet. For polymer-based resists thermally activated selective topography equilibration (TASTE) has been previously reported in the literature as the method to reduce the roughness of the blazed facet \cite{schleunitz_2011_taste,mccoy_2028_fabrication}. However, no studies were done so far for relevant line densities of such gratings in the soft X-ray energy range or for gratings with very shallow blaze angles nor for large patterned apertures. Thermal reflow allows to reduce the amount of imperfection on the facet, but it may also affect the shape of the grating profile like changing the blaze to anti-blaze ratio \cite{schleunitz_2011_taste}. Blazed profile gratings are used in analyzers and optical systems where the signal-to-noise ratio is a leading factor. Achieving high signals while reducing the impact of the diffuse scattering is a key goal for high-quality gratings. In the case of blazed gratings, the signal is larger when the ratio of blaze- to anti-blaze angle is as high as possible. Thus, controlling this step is critical to achieve a high signal-to-noise ratio.

Developing new methods, i.e. a sustainable fabrication process, requires metrology methods that can track these changes and give feedback on the process. EBL possesses all the requisites to produce a high-quality blazed-profile grating. It allows the patterning of large areas in a short time compared to the time consuming mechanical ruling and it is much more flexible in terms of shape-patterning. But to ultimately establish a reliable method for the fabrication of future smart gratings, a critical and systematic analysis of all the steps is needed. \\

We report on the development of a method for producing blazed profile gratings with very low facet roughness using EBL supported by highly precise metrological control of the process. For the development of the method a polymer-based resist (PMMA) was used. Different temperatures and heating times were tested during the thermal reflow phase, obtaining the optimal conditions for producing blazed profile gratings. A facet roughness close to the nominal roughness of Si wafer is obtained after etching the patterned structure into Silicon.  Atomic force microscopy (AFM) measurements as well as at-wavelength metrology were used to characterize the structures and to establish the technology.

\section{Method}
For the development of the lithography method a moderate line density of 600 lines/mm was chosen. Blazed gratings of that line density can be found in monochromators and analyzers in the soft X-ray energy range e.g., at the BESSY-II synchrotron \cite{maxymus_600lmm_2010,Schafers_journalSR_2026,belchem_600lmm_2022}. Additionally, the development of the lithography process had the emphasis to establish a highly reproducible and stable technology line for patterning blazed-profile structures into silicon. The latter would be an essential advantage compared to the classical mechanical ruling, where every 2nd to 3rd approach of ruling failed.

For the optimization of the process, it is inevitable to develop a systematic approach for controlling the different steps. Fig. \ref{fig:methods} a) shows a straightforward method for obtaining the nano-patterning of a grating processed into a substrate of single crystal Silicon using EBL. Additional steps, Fig. \ref{fig:methods} b) and c), were introduced to this approach to further reduce the facet roughness or have a better control of the apex region of the blazed profile grating. The e-beam writing process has been carried out at Helmholtz-Zentrum Berlin (HZB) as well as the ex-situ and at-wavelengths characterization. The transfer of the e-beam written pattern from resist into the wafer was done by means of Ar$^+$ ion beam etching at NOB Nano Optics Berlin GmbH (NOB).

\subsection{1. Manufacturing}

The process has been developed on 4-inch-Si wafers (100), under the consideration that it will be transferred to thicker substrate (up to 10 mm) in a future stage. Si-wafers are spin-coated with PMMA (671.02, \textit{Allresist GmbH}) with 2500 r.p.m. Higher spin-coating speeds, if possible, were avoided, because for a thick, heavy, rectangular shaped substrate the spin-coater speed is limited. After spin-coating, the wafers are baked in a convection oven at 180 °C. Unlike the more commonly used hotplate method, the convection oven helps minimizing the influence of varying ambient conditions, such as temperature or humidity, on a final structure.

\subsubsection{Electron beam lithography}
The e-beam system used for patterning is an EBPG5000 Plus from RAITH. EBL exposure parameters of 100 keV acceleration voltage and 10 nA beam current were used. 
For the production of blazed profile gratings different techniques for the electron beam exposure can be used. One method consists of tuning the dose per shot along the grating period, following the decay that corresponds to the blaze angle. Another method is multipass writing, which consists of varying the shots per area, increasing or decreasing the dose density, reaching a similar effect. Moreover, a combination of the two approaches is also possible, obtaining smoother transitions on the blazed profile at a cost of the exposure time. 

Here, we have divided the blazed angle into sections and adapted the doses along the blaze facet of the grating to obtain a gradient on PMMA that is then transferred to the Si-wafer by means of ion beam etching. The blazed facet is approximated by a finite number of steps, and the dose is changed accordingly to obtain the desired slope. The dose values  were varied between 50 $\mu$C cm$^{-2}$ and 250 $\mu$C cm$^{-2}$. 
To achieve this, we used a custom Python script to generate the GTX file, which was subsequently converted into the Raith EBPG pattern data format (GPF). 

After exposure, the samples are dipped in a bath of methyl isobutyl ketone (MIBK) (ALLRESIST, AR 600-56) for 60 s and then in a stopper (AR 600-60, ALLRESIST) for 10 s. After this process the blazed profile structures on PMMA are on the wafer.

\subsubsection*{Tempering}

Thermal annealing of 3D polymer-based structures has been already tested in the literature\cite{schleunitz_2011_taste}. However, the studies did not include large periods of the grating with very shallow blazed angles, and over large structured areas. By thermal annealing the blazed PMMA gratings, the facet roughness is reduced.\begin{wrapfigure}[26]{r}{0.6\linewidth}
    \includegraphics[width=\linewidth]{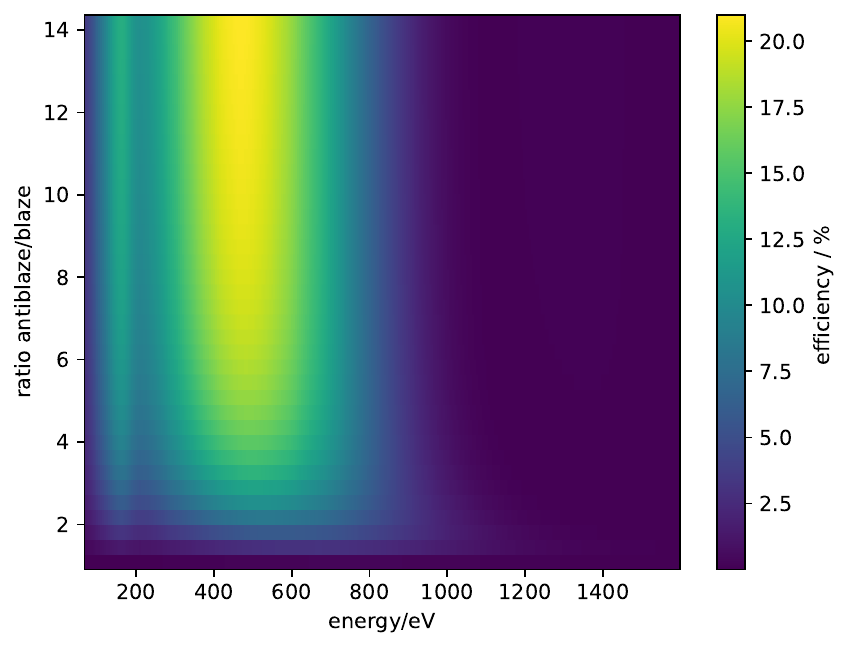}
    \caption{Efficiency calculation using REFLEC for the efficiency of a Silicon blazed grating with 600 lines/mm in the depicted energy range. Having a larger antiblaze to blaze ratio ensures better grating performance.}
    \label{fig:blaze_antiblaze}
\end{wrapfigure}
The samples were annealed using a convection oven, allowing for a uniform distribution of the hot air. The gratings were loaded and the temperature was ramped up to enable thermal reflow. They were held at the target temperature for a defined period. Both the reflow temperature and the annealing time were parameters under investigation, see Tab. \ref{tab:roughness}.

However, thermal reflow can also impact the shape of the apex region of the grating. This step is critical for the production of high-quality blazed gratings with a good blaze angle to antiblaze angle ratio and low facet roughness. Figure~\ref{fig:blaze_antiblaze} shows the dependency of the grating efficiency with the antiblaze-to-blaze ratio. The calculation was done for the soft X-ray energy region using REFLEC and considering 600 lines/mm gratings in Si, without coating. Having antiblaze angles as large as four times the blaze angles ensures a high quality grating.  Therefore, the development was focused on achieving good antiblaze to blaze angular ratios, which allow better signals, as well as the reduction of the facet roughness to reduce the noise.

\subsubsection*{Ion beam etching}

\begin{figure}
    \centering
    \includegraphics[width=0.7\linewidth]{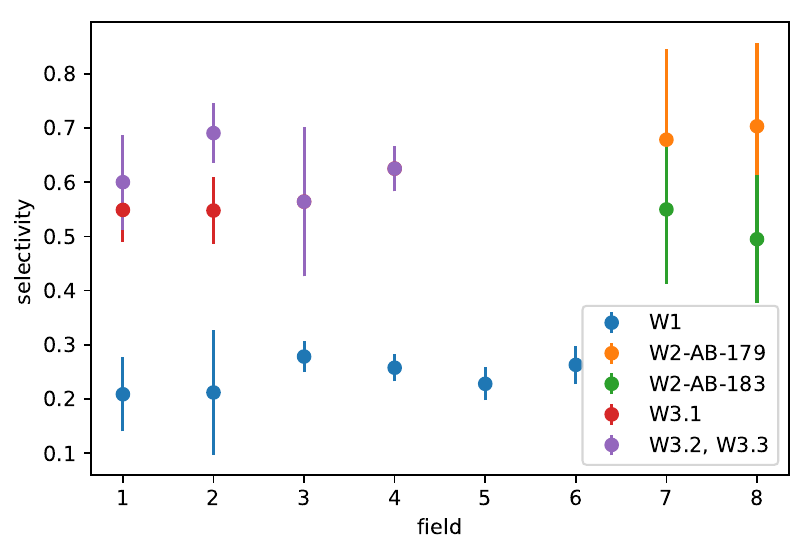}
    \caption{Selectivity rates of Si/PMMA for different set-ups. The different colors correspond to different wafers and test of the etching process.  }
    \label{fig:selectivity_ion}
\end{figure}

The structure produced on PMMA is then transferred into Si by ion beam etching at NOB Nano Optics Berlin GmbH. The etching machine of NOB allows to etch substrates with surfaces up to 200 mm x 200 mm and a thickness up to 30 mm. Etching depths from a few nm to 100 nm are possible. The substrates are moved during the etching process and different areas can be selected by an aperture above the sample, allowing to etch different depths in one process. Also depth gradients in one direction can be reached.  Figure \ref{fig:selectivity_ion} shows the selectivity rates of Si/PMMA for different sample sets. The values were calculated based on the measurements with AFM before and after etching. The colors on the graph indicate the same wafer, which was produced with different grating fields. The dispersion in the measurements corresponds to variation within the fields.

\subsection{2. Metrology}
Every step of the process was analyzed using appropriate measurement methods to establish a reproducible technique for patterning blazed-profile structures, highlighting the essential role of metrology. To this end, control of the various steps was performed using both ex-situ and at-wavelength tools, resulting in a systematic approach, see Fig.~\ref{fig:methods}.

\subsection{Ex-situ metrology}

Scanning force microscopy usually called Atomic force microscopy (AFM) was used to measure the facet roughness as well as the blazed profile. The atomic force microscopy measurements were done at HZB. AFM measurements were performed with a Nanosurf FlexAFM 100x100. The custom made AFM from Nanosurf can host samples of up to one meter in length and 100 mm thickness. The grating was measured under dynamic mode condition with a standard pyramidal shaped silicon probe (Tap 190 AL-G) with a nominal tip radius of < 10 nm. The inspected area is 15 $\mu m$ x 5 $\mu m$, with 4500 points in the longer axis and 500 lines for the short axis. 

These measurements allow to verify the best strategy to follow or whether to introduce further steps on the lithography process might be advisable. Systematically, several images were taken from the different fields and then analyzed using a python routine, developed in-house, and based on the software Gwyddion \cite{Necas_gwyddion_2012,Necas_2020_gwyddion_roughness}. The code was benchmarked with Gwyddion obtaining very good results (1 \% deviation) for this type of gratings and reducing the amount of processing time drastically. The presented values in this manuscript correspond to the extracted paramters and the standard deviation from all the measurements of the field. 

\subsection{At-wavelengths metrology}

The at-wavelength metrology was performed at the Optics Beamline at BESSY II \cite{Schafers_journalSR_2026, Sokolov_2018_efficient} during several measurement campaigns. This beamline is dedicated for accurate at wavelength metrology on sophisticated XUV and soft X-ray optical elements like reflective or transmission diffraction gratings, multilayered systems, reflection zone plates and etc. Higher order suppression system of the beamline provides spectral pure intensity in photon energy range from 12.5 eV up to 1850 eV is usable for quantitative measurements or reflection, diffraction or transmission. UHV 11-axis reflectometer can handle both small test samples and real-size optics up to 550 mm in length and up to 4 kg. It has goniometers for 360° in-plane rotation, for sample and detector stages. The sample stage is supported by an UHV tripod, which provides in-situ motion of 6 degrees of freedom for accurate alignment and mapping of the surface of the tested sample. Various apertures in front of the detectors allow to record the full beam reflected or diffracted from the sample as well as to resolve scattered intensity in the plane of the incident beam. 
The beam size in the sample position of (0.35 x 0.25) mm$^2$, vertical x horizontal, provides an optimal spot size on the sample and rather small vertical beam divergence of 0.5 mrad helps to keep beam from sample to detector with minimal distortions. 
The gratings were characterized by the collection of scattered and diffracted signals to obtain their efficiency and uniformity, as well as their dispersion pattern.

Measurements of the dispersion curves were performed at a fixed photon energy of 400 eV and an incidence angle of 4°. The detector arm was rotated to capture the scattered signal from both the zero and first diffraction orders, as well as the diffuse scattering in between. For the efficiency measurements, the incidence angle remained at 4°, while the first-order diffraction signal was recorded across a range of photon energies. The beamline energy was scanned from 60 eV to 700 eV. The chosen incidence angle ensured that the small test fields were not overilluminated, while still providing a sufficiently strong signal for reliable statistics.

\begin{figure}[!ht]
    \centering
    \includegraphics[width=0.8\linewidth]{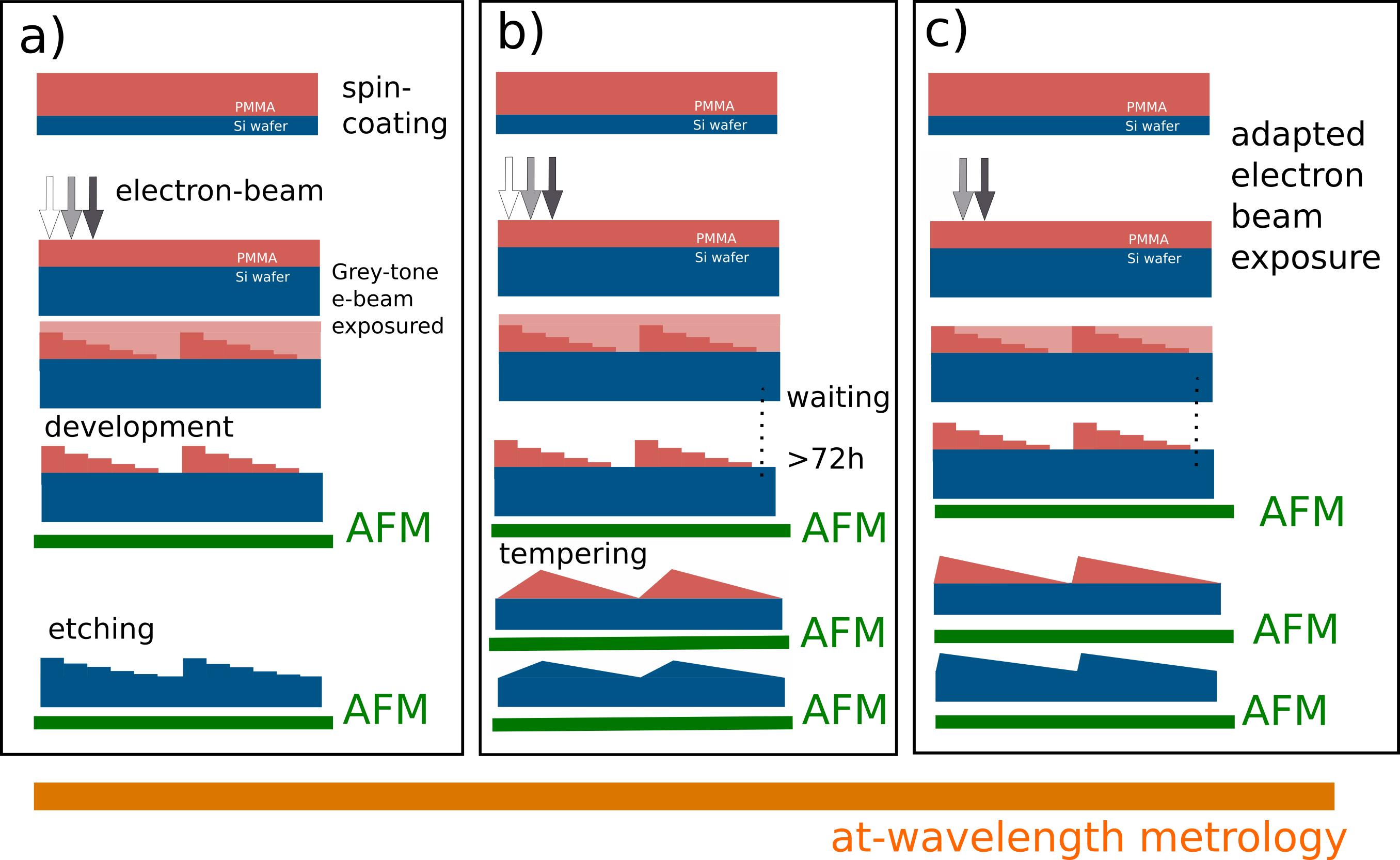}
    \caption{Manufacturing process and metrology methods used to control the processing of the gratings. a) shows the skeleton of the production of blazed profile gratings in a general way using  EBL and PMMA as positive resist. In b), tempering is included in the process to reduce the amount of roughness. At the end, see c), also waiting time between the exposure and developing was introduced as well as the adaptation of the writing process to better control the apex region of the gratings and counterbalance the effect of tempering. It should be noticed, that by using the shape \textit{gradient cube} for e-beam exposure the transition between the EBL exposure doses is smoother.  }
    \label{fig:methods}
\end{figure}

\section{Results and Discussion}

\subsection{Blazed profile fidelity}

\begin{wrapfigure}[16]{l}{.7\linewidth}
    \includegraphics[width=0.7\textwidth]{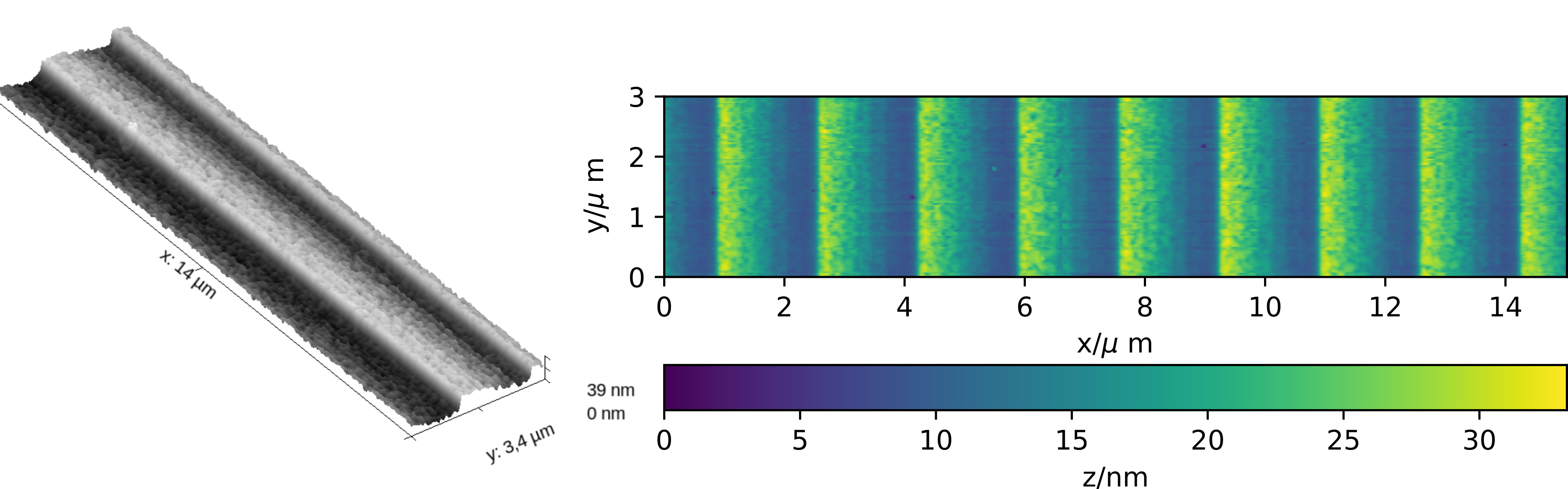}
    \caption{Facet roughness up to 1.5 nm is obtained for the structures on resist (PMMA) before they are tempered or transferred into the substrate. In this case the blazed profile was approximated by 7 steps.}
    \label{fig:pre-roughness}
\end{wrapfigure}

To fabricate the blaze-profile grating, elementary shape primitives such as \textit{path} or \textit{trapezoid} shapes in the EBPG were utilized and tested. To analyze the influence of the steps approximating the blazed profile, a set of samples was produced, considering 7, 17 and 119 (using \textit{path}) dose steps by a beam step size (BSS) of 0.014 $\mu m$.  According to the results, small variations to the blaze angle were observed (less than 5 \%) between the different approaches. The field whose facet is approximated by fewer steps presented shallower blazed angles. We attribute this effect to the large terraces used to approximate the facet gradient, together with the proximity effect, which lead to less pattern fidelity. The terraces are wider when the number of steps is smaller. The applied average doses are slightly different and would result in a different resist height after development. If not indicated otherwise, the shown gratings correspond to blazed facets approximated by 17 steps.  Using this number of steps and the process shown in Fig. \ref{fig:methods} a), facet roughness as low as 1.5 nm was achieved on PMMA \ref{fig:pre-roughness}, going down to 1.09 nm after etching, still larger than nominal values for Si. Although increasing the number of steps reduced the facet roughness slightly, the improvement was not sufficient to eliminate the need for tempering. 

Following exposure, the samples were developed at different times after exposure. As previously reported~\cite{MORTELMANS_2020_grayscale}, the contrast curve of PMMA is sensitive to the delay between exposure and development. In this context, such variations would result in deviations in the resulting blaze angles. In our test it was observed macroscopically that the overall homogeneity of the fields and the facet roughness of the blazed gratings were affected. This dependency on the time span between exposure and development is shown in Tab. \ref{tab:waiting}. The test was done using six different fields on the same wafer and several points were measured using AFM, at the center and edges of the field. After the exposure the wafer was taken out of the system and kept in ambient conditions in a clean room. The roughness value corresponds to the standard deviation over all measured points within the field. And, for the analysis, any deviation from a perfect facet line is considered. The values for the blazed angle and height correspond to the average and standard deviation of all analyzed data for each field. Thus, the non-uniformity of the blazed profile grating can be seen on the standard deviation of the height. It should be noted that the absolute values are not statistically significant and depend as well on ambient conditions, however, these observations were consistent.

\setlength{\tabcolsep}{10pt} 
\renewcommand{\arraystretch}{1.5} 
\begin{table}[!ht]
\centering
\begin{tabular}{c c c c }
\hline
 waiting time / h & blazed angle / ° & height / nm &   roughness / nm \\
\hline
          6 < t < 12 & 1.51 $\pm$ 0.08  & 37.5 $\pm$ 1.6 & 1.8 \\ 
          14 < t < 24 & 1.43 $\pm$ 0.08  & 31.6 $\pm$ 0.6 & 1.6  \\
          > 72 & 1.23 $\pm$ 0.06 & 31.5 $\pm$ 0.5 & 1.4 \\
\hline
\end{tabular}
\caption{Facet roughness on PMMA measured with AFM depending on the waiting time between exposure and development. Please notice that the difference between the blazed angles for similar heights is due to different antiblazed angles. }
\label{tab:waiting}
\end{table}

\subsection{Facet roughness and apex angle}

  \begin{wrapfigure}[16]{l}{.6\linewidth}
    \includegraphics[width=0.6\textwidth]{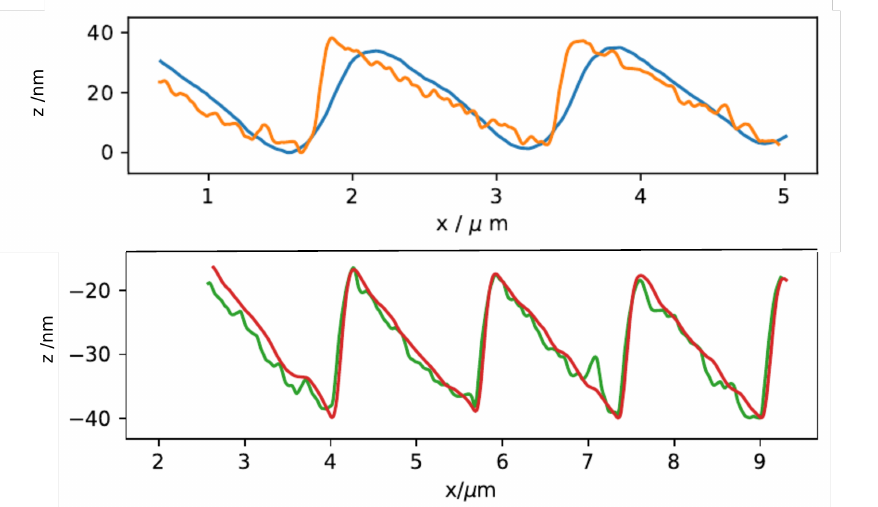}
    \caption{AFM profiles of the thermal annealed fields: (top) at 120°C following the steps in Fig. \ref{fig:methods}b) and at the bottom, after the adaptation of the writing procedure following the steps in Fig.\ref{fig:methods} c), and tempering the sample for 15 min at 125 °C.}
    \label{fig:adapted_EBL}
\end{wrapfigure}
\begin{figure}[hb]
    \includegraphics[width=1.\linewidth]{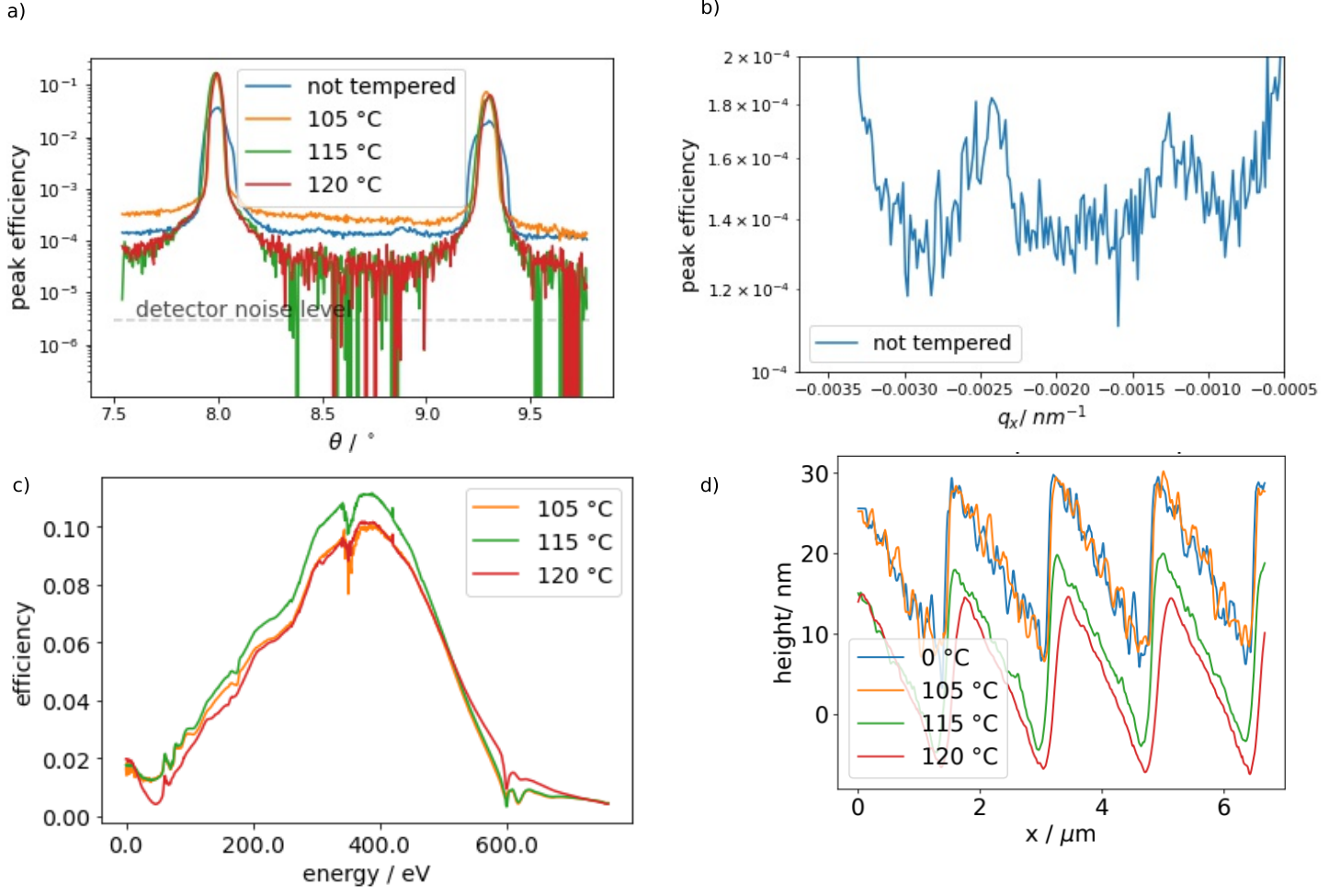}
    \caption{a) Dispersion curves of the etched gratings without being annealed (blue) and after tempering for three different temperatures during 120 min. The grating that was not tempered show the contribution of higher periodicities, produced by the 4.5 $\mu m$ subfield of the e-beam writer (b). This fades out when the sample is tempered.  c) The at-wavelengths efficiency of the gratings after annealing shows that tempering influences the processed grating and impacts the efficiency. d) The profile of each of the gratings before etching is shown.}
    \label{fig:at-wavelenths_tempering}
\end{figure}

  Thermal annealing of the PMMA gratings lead to a great reduction of the roughness, see table \ref{tab:roughness}. It shows the correlation between temperature and the reduction of the roughness. After etching those values would decrease further. The impact on the antiblaze to blaze ratio is also observed in this process, affecting the shape. Thus, reducing the roughness does not always mean that the efficiency would simultaneously increase. The gratings were tested at-wavelengths after being etched. Figure \ref{fig:at-wavelenths_tempering} shows the impact of tempering on the dispersion pattern of the grating as well as the efficiency at-wavelengths. The diffuse scattering is reduced when the grating is annealed, see \ref{fig:at-wavelenths_tempering} a). The superperiodicity that is visible for the non-tempered sample, as well as for the lowest temperature, corresponds to the used sub-field of the e-beam writer of 4.5 $\mu m$ (see Fig. \ref{fig:at-wavelenths_tempering} b)). This contribution becomes negligible after increasing the annealing temperature. This effect can also be minimized by optimizing the writing process during e-beam writing, i.e. optimizing the direction and ordering of the exposures to minimize this impact. 

\begin{wrapfigure}[20]{r}{.5\linewidth}
    \includegraphics[width=0.4\textwidth]{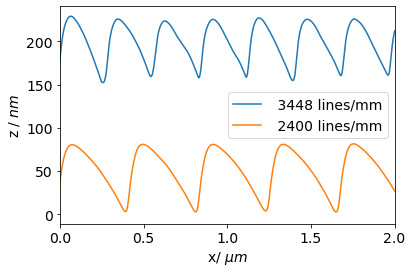}
    \caption{AFM profiles on PMMA of the high line density test using \textit{path}.}
    \label{fig:hld}
\end{wrapfigure} However, lowering the facet roughness does not directly involve an increase on the grating efficiency, see Fig.\ref{fig:at-wavelenths_tempering} c). The deformation of the shape through thermal annealing influences negatively the efficiency. This behavior is attributed to the loss of sharpness in the apex region following annealing, and the variation of the blaze and antiblaze angles, as it was predicted in Fig.\ref{fig:blaze_antiblaze}.   Lines from the AFM measurements, shown in Fig.\ref{fig:at-wavelenths_tempering} d), illustrate the impact that tempering has on the apex region as well as in the facet roughness. The corresponding quantitative values are summarized in Table \ref{tab:roughness}. The thermal reflow induced by annealing smooths the exposed area but also alters the grating profile, meaning that this reshaping must be accounted for in the fabrication process to achieve a grating with a specific blaze angle. Different blaze angles will shift the maxima of the efficiency on the energy axis. For this reason, the non-tempered sample was left out of the efficiency comparison.\\

In a further analysis, the writing method was adapted to leave the apex region without being exposed, as it was previously tested by Schleunitz et al. \cite{Schleunitz_2010_dose-variation}. This led to a better antiblaze to blaze angular ratios, see Tab. \ref{tab:roughness}. Samples 4 to 7 were tempered using the same temperature, but for different times. The samples indicated with * correspond to the ones, where the exposure left out the apex regions of the grating. By doing that, the apex region remains sharp and pattern fidelity increases. However, it was observed that in order to be significant, at least 5 \% of the area at each period should not be exposed to any dose. Fig. \ref{fig:adapted_EBL} shows the comparison of the fields before and after tempering. By only leaving one spot (1 \%) of the apex region without a dose (upper image) the desired effect was not totally achieved. This might be critical for very small periods, where the number of beam steps is reduced.

\begin{table}[!ht]
\centering
\begin{tabular}{c c c c c c}
\hline
Sample & Initial Roughness / nm & Temperature / $^\circ$C & Time / min  & Roughness After / nm & Antiblaze-to-Blaze\\
\hline
0 & 1.5 & - & - & - & 12 $\pm$ 5 \\
1 & 1.5 & 105 & 120 & 1.3 & 9 $\pm$ 4\\
2 & 1.6 & 115 & 120 & 0.7 & 6.0 $\pm$ 0.7 \\
3 & 1.5 & 120 & 120 & 0.4 & 3.8 $\pm$ 0.3\\
\hline
4 & 1.1 & 125 & 15  & 0.6 & 5.6 $\pm$ 0.5\\
5* & 1.1 & 125 & 15  & 0.6 & 6.6 $\pm$ 0.4\\
6 & 1.7 & 125 & 30  & 0.5 &  5.9 $\pm$ 0.4\\
7* & 1.7 & 125 & 30  & 0.5 & 8.0 $\pm$ 0.5\\
\hline
\end{tabular}
\caption{AFM analysis of the gratings in PMMA before and after thermal treatment at various temperatures and durations. * These samples were produced by adapting the exposure, allowing the apex region to not receive any dose. The impact of the adapted exposure can be seen in the AFM profiles, Fig. \ref{fig:adapted_EBL}. }
\label{tab:roughness}
\end{table}

This redefined writing process, leaving out the apex region from the exposure, was tested for 2400 lines/mm as well as for the extreme case of a grating with 290 nm periodicity (3448 lines/mm). In this latter case, the 5 \% of the total doses to exposed a period corresponds to a unique shot. Fig. \ref{fig:hld} shows the AFM measurement for both gratings. For the extreme case the antiblaze to blaze ratio is around 2.5. The case of 2400 lines/mm gratings shows a better angular ratio, that was still better conserved after tempering. The angular ratio in this case is still larger than 4.  However in these cases, other approaches might be tested that involve changing the overall parameters of the exposure, such as BSS, or vary the dose density by using multipass.

\begin{figure}[h!]
    
    \includegraphics[width=1.\linewidth]{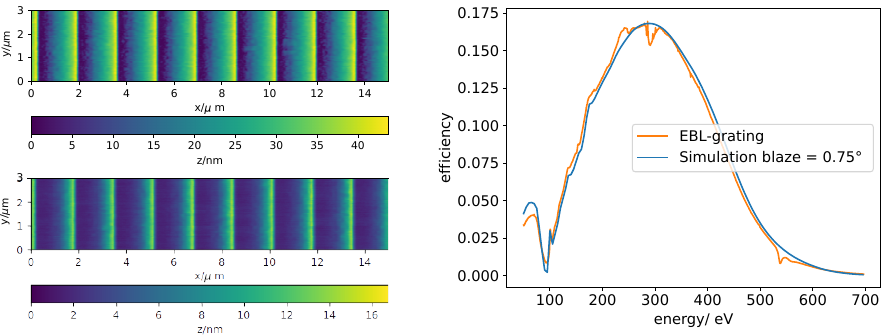}
    \caption{AFM measurements of the grating before etching (top) and after etching (bottom). The facet roughness was 0.74 and 0.39 nm, respectively. On the right is the measured efficiency at-wavelengths.}
    \label{fig:best-grating}
\end{figure}

Figure \ref{fig:best-grating} shows the AFM  profile as well as the efficiency curve of a grating produced following the above mentioned considerations. See Fig. \ref{fig:methods} c) for a summary of those steps. By doing so, the roughness of the facet after etching in Si is 0.39 nm and the antiblaze to blaze ratio $6.5 \pm 0.5 $. The expected efficiency was calculated using REFLEC for a grating with a blaze angle of 0.75° and an ideal antiblaze angle. The calculation and the measurement shows a very good agreement and the variations might be due to small variations of the grating profile. The AFM measurement, although performed at several locations, is still a local method, while the at-wavelengths measurements gives a global picture of how this type of grating would perform under real illumination. The simulated ideal profile does not consider any contamination from carbon or oxygen, as the measured efficiency of the grating does. Usually, this grating is used after coating it with a high-reflective material or a multilayer, depending on the energy region, and increasing its final efficiency.

Lastly, a new shape primitive implemented by RAITH, named \textit{gradient cube}, was tested as well. It allows the variation of the dose along a defined axis. By corresponding this axis with the grating period, it is possible to create a grating with a specific blazed profile angle. This is achieved by adjusting the bottom and top doses of each facet, as determined by the contrast curve and in between the dose is varied for each shot on the facet. This is similar to the procedure of using \textit{path}, and so were the results. 
However, \textit{gradient CUBE} simplified the process of writing blazed profile gratings using E-beam exposure and reduces the size of the created file, thus improving its readability and integration. This is crucial for more sophisticated e-beam exposures, such as those needed for variable line space gratings over large aperture areas.

\section{Conclusion}
In this study, we show the feasibility of producing highly precise blazed gratings by means of EBL. Moreover, we show that the availability of dedicated metrology is a key factor in decision making in nanopatterning technology. The steps introduced in the process, such as waiting and tempering, leverage the impact of the roughness on the facet as well as improve the overall shape of the blazed profile gratings. Why find that by leaving 5\% of the grating, corresponding to the apex region of the grating, unexposed, higher ratios of antiblaze of blaze angles can be obtained. However, this is challenged by very small gratings period.

The patterning time of the EBL system for a 600 lines/mm grating would take the similar amount of time as mechanical ruling, considering as well the equilibration time inside the EB writer. However, the EBL writing time does not increase significantly with the line density as it is the case with mechanical ruling. The latter is a major advantage when producing large patterned apertures of significantly higher line density.  Additionally, EBL is very versatile and it allows to control the resist exposure and adapt it locally, which makes this method perfectly suitable to the production of variable line space (VLS) gratings.

Follow-up studies on benchmarking EBL to other existing lithography methods, can finally help to establish EBL as patterning technology for blazed profile gratings in small and medium sized apertures. Up-comming work should also cover investigations on modifying the concept of reflection zone plates (RZP) \cite{Erko:14,Braig_14}towards RZP`s with blazed groove profile instead of lamenar. However, the thermal load on the substrate of these optical elements in FELs and synchrotron radiation facilities is still a topic to be addressed.

 \begin{acknowledgments}
 
This research was supported by the European Union’s Horizon 2020 research and innovation program under grant agreement No. 101004728 (LEAPS-INNOV). The authors would like to thank all our partners within the LEAPS-INNOV WP4-NeXtgrating cooperation for very useful discussions and exchange of ideas.  AFH, AS and FS are very grateful to Alexei Erko (formerly at HZB) for many years of coworking, sharing his ideas to us and profound discussions on the topics shown in this publication. 

\end{acknowledgments}

\bibliography{references}

\end{document}